\documentclass[prl,twocolumn,superscriptaddress,showpacs]{revtex4}
\usepackage{latexsym}
\usepackage{graphicx}
\def\H{{\cal H}}

\def\ket#1{| #1 \rangle}

\newif\ifpdflatex\pdflatextrue
\makeatletter\@ifundefined{pdfoutput}{\pdflatexfalse}\makeatother
\def\myincludegraphics[#1]#2#3{%
\ifpdflatex \includegraphics[#1]{#2}
\else       \includegraphics[#1]{#3}
\fi}
\begin{document}
\title{Implementation of controlled multi-qudit operations for a solid-state quantum computer based on charge qudits}
\author{S.~G.\ Schirmer}
\email{sgs29@cam.ac.uk}
\affiliation{Department of Applied Mathematics and Theoretical Physics,
University of Cambridge, Wilberforce Road, Cambridge CB3 0WA, UK}
\affiliation{Department of Engineering, University of Cambridge,
Trumpington Street, Cambridge, CB2 1PZ, UK}
\author{Andrew~D.\ Greentree}
\email{a.d.greentree@unsw.edu.au} 
\affiliation{Centre for Quantum Computer Technology, School of Physics, 
University of New South Wales, Sydney, NSW 2052, Australia}
\author{D.~K.~L.\ Oi}
\email{dklo2@cam.ac.uk}
\affiliation{Department of Applied Mathematics and Theoretical Physics,
University of Cambridge, Wilberforce Road, Cambridge CB3 0WA, UK}
\date{\today}
\begin{abstract}
We consider a mechanism to generate controllable qudit-qudit interactions in
a charge-position paradigm for a quantum computer, through the use of auxiliary
states.  By controlling the tunneling rates onto these auxiliaries from the
qudits proper, we can controllably switch the entangling operations.  We
consider a practical architecture in which to realize such a computer and
examine the associated Hilbert space dimension.
\end{abstract}
\maketitle

Quantum computing~\cite{bib:NielsenBook} has been identified as an important
field recently, and significant work is being undertaken to find suitable
systems in which to observe scalable, coherent interactions.  Recent advances
in semi-conductor technology have opened up new possibilities for silicon-based
solid-state realizations of quantum computers, which are highly attractive due
to their compatibility with conventional Silicon metal-oxide-semiconductor
technology~\cite{bib:ClarkReview2003}. All scalable quantum computing proposals
require some particle that can be placed into a superposition of several 
distinguishable quantum states, and for which particle-particle entanglement
can be achieved.  The Hilbert space dimension of an $N$ particle quantum 
register obtained by entangling all the particles simultaneously has been 
shown to be a good measure for the power of a quantum 
computer~\cite{bib:Blume-Kohout2002}.

One model for a scalable quantum computer is based on localization of an
electronic charge in potential wells, the so-called charge-qubit approach.
This has been discussed by Ekert~\cite{bib:Ekert1995} and appears in various
flavors in the literature, see for example Refs \cite{bib:EkertRMP1996, 
bib:BarencoPRL1995}.  A generic problem with charge-based quantum computing
schemes is that, although it is relatively easy to generate single particle
operations, i.e., to realize the Hadamard transform gate, it is usually 
quite difficult to obtain \emph{controllable} particle-particle interactions.
The mechanism normally suggested is the Coulomb interaction, which has the 
problem of being difficult to turn off.  It may be possible to perform 
universal operations with an ``always on'' interaction, and schemes for 
realizing global operations in this setting have been considered by 
Benjamin~\cite{bib:BenjaminPRL2003} and Pachos and
Vedral~\cite{bib:PachosPreprint2003}.  However, for practical purposes it
seems highly desirable to have controllable multi-particle interactions.
Here we suggest ways to achieve this by making use of local tunneling to
auxiliary states, which has the effect of switching the Coulomb action by
changing the effective distance between components of the electronic
wavefunction.

The basic schemes we propose are in principle not limited to a
particular charge-based implementation, and may in fact be useful
for a variety of charge-based proposals including, for instance,
Cooper-pair box schemes~\cite{bib:MakhlinNature1999}.  However,
for concreteness we consider a generalization of a specific recent
proposal by Hollenberg \emph{et al.}~\cite{bib:Hollenberg2003}
where the confining potentials are obtained from individual
Phosphorus implants in a Silicon matrix.  We envisage this being
fabricated via a ``bottom up'' approach to nanofabrication such as
has been recently realized~\cite{bib:OBrienPRB2001}. Hollenberg's
charge proposal deserves particular consideration in our opinion
since it takes advantage of existing fabrication technologies and
medium-scale realizations of such an architecture appear to be
within experimental reach in the near future. Although the
original proposal involved charge qubits, we shall consider a
generalization to qudits (i.e., systems whose single particle
Hilbert space has dimension $D\ge 2$), and concentrate on qutrit
structures ($D=3$) as they have been shown to optimize the total
Hilbert space dimension of the composite 
system~\cite{bib:GreentreePrePrint2003}.  We propose concrete scalable
architectures that permit efficient, controlled multi-particle
gates in this setting, which is essential for the operation of
quantum error-correction algorithms.

\begin{figure}
\myincludegraphics[height=2.4in]{figures/pdf/potential-well.pdf}{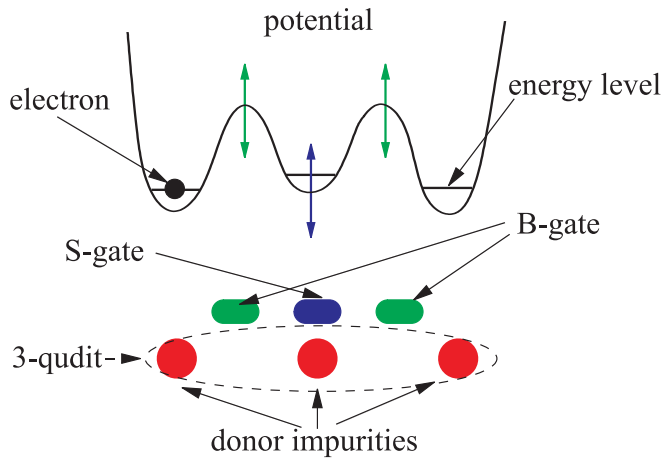}
\caption{Schematic of potential created by three donor impurities forming
         a charge qutrit with ground state energy levels of the shared
         electron, shown here localized in the left well, as well as a
         S-gate above the center dot $2$ and two B-gates between dots
         $1$, $2$ and $2$, $3$, respectively.} \label{figure1}
\end{figure}

The donor impurities in the Hollenberg \emph{et al.} scheme can be
regarded as quantum dots that create an electric potential with
$D$ local minima located at the sites of the donor impurities,
which confines the shared electron as shown in Fig.~\ref{figure1}.  
The $D$ ground states $\ket{d}$ for $d=1,\ldots,D$, corresponding
to the various localizations of the
electron at the donor impurities, form a basis for the Hilbert
space $\H$ of a single charge qudit.  The height of the potential
barrier between a pair of adjacent quantum dots belonging to the
same qudit can be manipulated by applying an external electric
potential via surface electrodes located between the two sites. By
adjusting the voltages applied we can manipulate the height of the
potential barrier and therefore the rate of tunneling between
adjacent sites.  We follow the notation of Ref~\cite{bib:Hollenberg2003} 
and refer to these surface electrodes as barrier or $B$-gates.  
Quantum tunneling through the potential barriers leads to the 
creation of coherent superposition states of the localized charge
qudit states $\ket{d}$.  Furthermore, by applying electric 
potentials to surface electrodes located directly above the donor
impurities we can create asymmetries in the potential, and thus 
change the ground state energy $E_d$ of state $\ket{d}$, introducing
local energy shifts.  We shall call these electrodes (energy) shift 
gates, or $S$-gates.  It can be shown that by combining $S$-gates 
and $B$-gates, arbitrary single qudit operations (local unitary 
operations) can be performed.

\begin{figure}
\myincludegraphics[height=2.4in]{figures/pdf/scheme0.pdf}{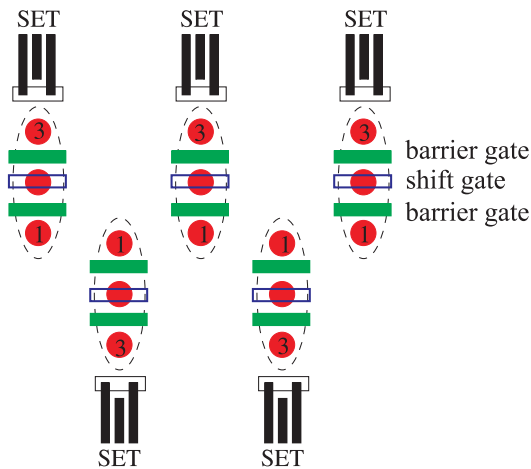}
\caption{Scalable arrangement of quantum dots partitioned into qutrits with
        fixed, permanent Coulomb interaction between $\ket{1}$-states of all
        adjacent qudits.}
\label{figure2}
\end{figure}

One of the main advantages of the proposed charge qudit
architecture compared to the original Kane proposal~\cite{bib:Kane1998} 
using the nuclear spins of the donor atoms as qubits is the possibility
of \emph{easy readout} of quantum information via single electron 
transistors~\cite{bib:GrabertNATO1992}, the suitability of which has 
already been shown for quantum computing applications when run in RF 
mode~\cite{bib:RFSET}. Moreover, charge qudit quantum logic operations
can be implemented using voltage gates only, without the need for
additional radio frequency control pulses. Besides reducing the
operational complexity, this should largely avoid problems of
non-selective excitation of multiple sites, which may complicate
the implementation of selective single and multi-qubit quantum
logic operations in the original Kane proposal.

One of the main disadvantages of charge qudits compared to nuclear
spin qubits is their much shorter coherence lifetime.  However,
local operations can be performed much faster for charge qudits
than for nuclear spin qubits, on the order of $10^{-11}$ seconds,
and estimates suggest that the decoherence lifetime of a charge
qudit should be much longer than that, at least on the order of 
$10$ ns~\cite{bib:Hollenberg2003, bib:Barrett2003}.  It also seems 
likely that advances in technology and control will lead to further
reductions of the gate operation time, and that the coherence 
lifetime might be increased though decoherence control measures.  
For instance, continuous quantum error correction
by means of quantum feedback control has recently been shown to be
able to reduce spontaneous emission errors \cite{bib:Ahn2003}, and
similar control techniques might be able to reduce decoherence.

\begin{figure}
\myincludegraphics[width=0.45\textwidth]{figures/pdf/scheme1a.pdf}{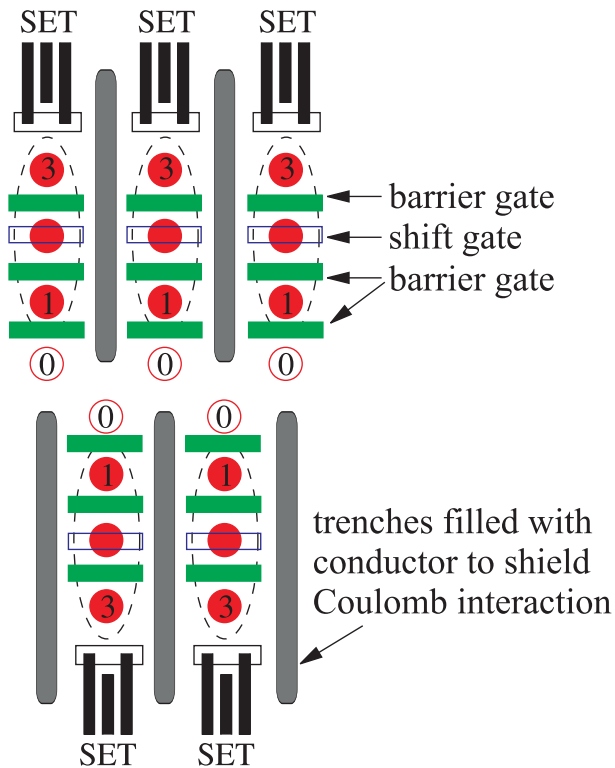}
\caption{Scalable arrangement of quantum dots partitioned into qutrits with
        auxiliary quantum dots to mediate the Coulomb interaction.}
\label{figure3a}
\end{figure}

We present our proposal for a scalable charge-qudit quantum
computer in Fig.~\ref{figure3a}.  As mentioned above, we
differentiate it from previous models by the use of (a) qutrits to
optimise the Hilbert-space dimensionality, and (b) auxillary dots
to effectively switch the Coulomb interaction to mediate
particle-particle interactions.  The qudits are arranged
vertically, alternatingly above and below a center row of
auxiliary quantum dots which mediate interactions between multiple
adjacent qudits. The alternating arrangement of the qudits above
and below the row of auxiliary quantum dots reduces unwanted
Coulomb interactions between quantum dots belonging to different
qudits.  The efficiency of this shielding effect can be further
enhanced by adding ``trenches'' filled with a conducting material
between adjacent qudits as shown in the figure.  Since the donor
impurities are buried, realizing such shielding trenches would
require a 3D structure. One way to realize such a structure would
be to metalize a sheet of delta-doped Silicon, which could be
achieved with a bottom-up strategy~\cite{bib:OBrienPRB2001}. 
The staggering of the qudits further reduces unwanted interactions
between diagonally facing qudits.  The
efficiency of multi-qudit operations might be enhanced by
embedding the auxiliary quantum dots into a high-permittivity
material to facilitate the Coulomb interaction. This would require
that the silicon substrate be replaced by another material in the
region occupied by the auxiliary quantum dots.  This would be
difficult to realize given current technology, however it may well
become feasible with further advances in fabrication technology.
The high permittivity material would reduce the time needed to
implement multi-qudit gates, increasing the number of such
operations that can be performed within the coherence lifetime of
the qudit register, potentially improving the performance of a
working quantum device. 

\begin{figure}
\myincludegraphics[width=0.45\textwidth]{figures/pdf/scheme1b.pdf}{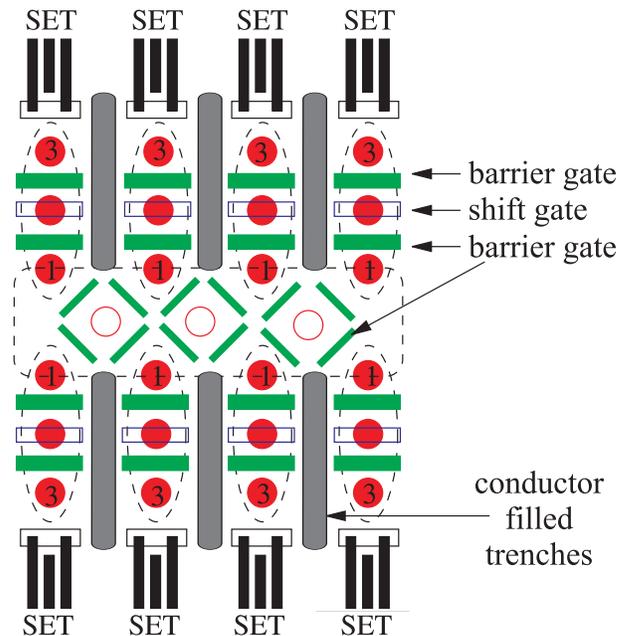}
\caption{Scalable arrangement of quantum dots partitioned into qutrits with
        horizontally offset auxiliary quantum dots to mediate the Coulomb
        interaction.}
\label{figure3b}
\end{figure}

To activate the Coulomb interaction between two or more adjacent qudits, we
coherently transfer the populations of state $\ket{1}$ of the corresponding
qudits to the auxiliary state $\ket{0}$ by \emph{simultaneously} lowering
the height of the potential barriers between the corresponding quantum dots
by applying suitable gate voltages to the auxiliary tunneling gates.  Then
we restore the potential barriers and let the Coulomb interaction between
the auxiliary quantum dots create the necessary controlled phase gate.  After
a certain predetermined time, the populations of the auxiliary quantum dots
are coherently transferred back to the original qudit states by simultaneously
lowering the corresponding auxiliary potential barriers, thus switching off
the Coulomb interaction.  The details of how to implement a sufficient set
of elementary gates required for universal quantum computation and efficient
qudit error correction will be discussed in a forthcoming, more detailed paper.

Note that this arrangement of the qudits is scalable and permits not only
the implementation of arbitrary two-qudit gates between any pair of adjacent
qudits by combining controlled phase gates with single qudit operations, but,
in principle, the implementation of arbitrary gates involving a set of $k\le N$
adjacent qudits.  Furthermore, since the effects of the control fields (voltage
gates) are strictly spatially confined, unlike for radio-frequency control
fields for instance, it should be easily possible to implement multi-qudit
interactions on disjoint subsets of qudits simultaneously.

One drawback of the proposed architecture is that it requires a comparatively
large number of auxiliary sites to mediate the interactions.  In principle, it
is possible to reduce the number of auxiliary quantum dots required by modifying
the architecture as shown in Fig.~\ref{figure3b}, for instance.  In this
modified arrangement the auxiliary sites are horizontally offset such that each
auxiliary site is shared by four qudits.  This reduces the number of auxiliary
sites required by half.  However, a potential drawback of this architecture is
that, while it should be easy to implement four-qudit gates, it would appear to
be difficult to implement controlled \emph{two}-qudit interactions since the
Coulomb interaction will always affect all four qudits surrounding an auxiliary
quantum dot.

\begin{figure}
\myincludegraphics[width=3in]{figures/pdf/Hilbert-dim.pdf}{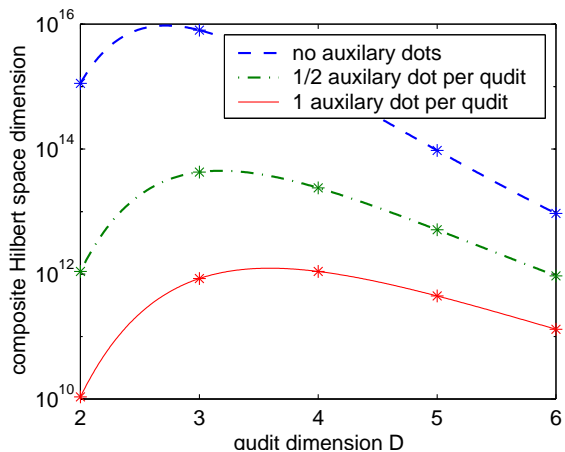}
\caption{Hilbert space dimension of a qudit register comprised of a fixed
number of quantum dots as a function of qudit size for the architectures
discussed.  For concreteness the values on the vertical axis correspond
to a total number of $K=100$ donor sites but the number of donor sites
does not change the graph qualitatively.}
\label{figure4}
\end{figure}

Finally, we consider the effect of the auxiliary quantum dots on the Hilbert
space dimension of the register.  As mentioned earlier, in absence of auxiliary
quantum dots, partitioning into qutrits will always optimize the Hilbert space
dimension of the quantum register.  Clearly, given any fixed number of quantum
dots, the need for auxiliary quantum dots reduces the dimension of the Hilbert
space of the quantum register.  If one auxiliary site is needed for each qudit,
as in the architecture shown in Fig.~\ref{figure3a} for instance, then the
dimension of the Hilbert space of a register of $K$ quantum dots partitioned
into $D$-qudits will be $D^{K/(D+1)}$ instead of $D^{K/D}$.  If only a single
auxiliary site for every two qudits is needed as in Fig.~\ref{figure3b} then
the Hilbert space dimension of the register will be $D^{K/(D+0.5)}$ instead.
Fig.~\ref{figure4} shows the Hilbert space dimension of a qudit register
comprised of a fixed number of quantum dots as a function of qudit size for
the three architectures discussed.  The graph clearly shows that partitioning
into qutrits maximizes the Hilbert space dimension of the quantum register if
either no auxiliary quantum dots or only a single dot per pair of qudits is
required.  If one auxiliary dot for each qudit is used then partitioning into
four-qudits increases the Hilbert space dimension slightly.  However, this
increase is rather small compared to the increase from qubits to qutrits and
may be offset by other considerations such as the increased complexity of
single qudit gates or the number of measurements required to extract a
comparable amount of quantum information for $D>3$.

In summary, we have proposed concrete scalable architectures for a charge
qudit quantum computer that allow the efficient implementation of controlled
multi-qudit gates by making use of auxiliary sites to mediate multi-qudit
interactions.  Our emphasis has been on the implementation of such schemes
for a practical realization of a charge qudit quantum computer based on donor
impurities embedded in a Silicon matrix, but they are not fundamentally
limited to this specific case.

SGS and DKLO are supported by the Cambridge-MIT Institute project on quantum
information.  ADG is supported by the Australian Research Council. \\


\end{document}